\documentclass[sigconf]{acmart}

\usepackage{booktabs}
\usepackage{algorithm}
\usepackage{algpseudocode}
\usepackage{bbold}
\usepackage{amsmath}

\newcommand{\todo}[1]{\textcolor{red}{#1}}

\newcommand{\sz}[1]{\textcolor{blue}{#1}}

\newtheoremstyle{mydef}
{2ex}
{2ex}
{\itshape}
{}
{\scshape}
{: }
{0.5em}
{}
\theoremstyle{mydef}
\newtheorem{mydef}{Definition}

\newcommand{\miniskip}{\vspace*{-.4\baselineskip}}
\newcommand{\shrink}{\vspace*{-.9\baselineskip}}

\begin{document}
\copyrightyear{2017}
\acmYear{2017}
\setcopyright{acmcopyright} 
\acmConference{SIGIR'17}{}{August 07--11, 2017, Shinjuku, Tokyo, Japan.} \acmPrice{15.00} 
\acmDOI{http://dx.doi.org/10.1145/3077136.3080796}
\acmISBN{978-1-4503-4887-4/17/08}

\title{EntiTables: Smart Assistance for Entity-Focused Tables}
\fancyhead{}

\author{Shuo Zhang}
\affiliation{%
  \institution{University of Stavanger}
}
\email{shuo.zhang@uis.no}

\author{Krisztian Balog}
\affiliation{%
  \institution{University of Stavanger}
}
\email{krisztian.balog@uis.no}

\begin{abstract}
Tables are among the most powerful and practical tools for organizing and working with data.  Our motivation is to equip spreadsheet programs with smart assistance capabilities.
We concentrate on one particular family of tables, namely, tables with an entity focus.  We introduce and focus on two specific tasks: populating rows with additional instances (entities) and populating columns with new headings.
We develop generative probabilistic models for both tasks.  For estimating the components of these models, we consider a knowledge base as well as a large table corpus.  
Our experimental evaluation simulates the various stages of the user entering content into an actual table.  A detailed analysis of the results shows that the models' components are complimentary and that our methods outperform existing approaches from the literature.
\end{abstract}



\maketitle

\section{Introduction}

Tables are one of the most effective and widely used tools for organizing and working with data.  Spreadsheet programs are among the most commonly used desktop applications, both in business environments and in personal use, because of their ease of use and flexibility.
The overall objective of this study is to develop an intelligent personal assistant that can offer smart assistance for people working with tables.  It may be imagined as the infamous Office Clippy, albeit we prefer it to be less obtrusive. 
This study represents the first step towards this ambitious endeavor.

\begin{figure}[t]
   \centering
   \includegraphics[width=7.5cm,height=5.0cm]{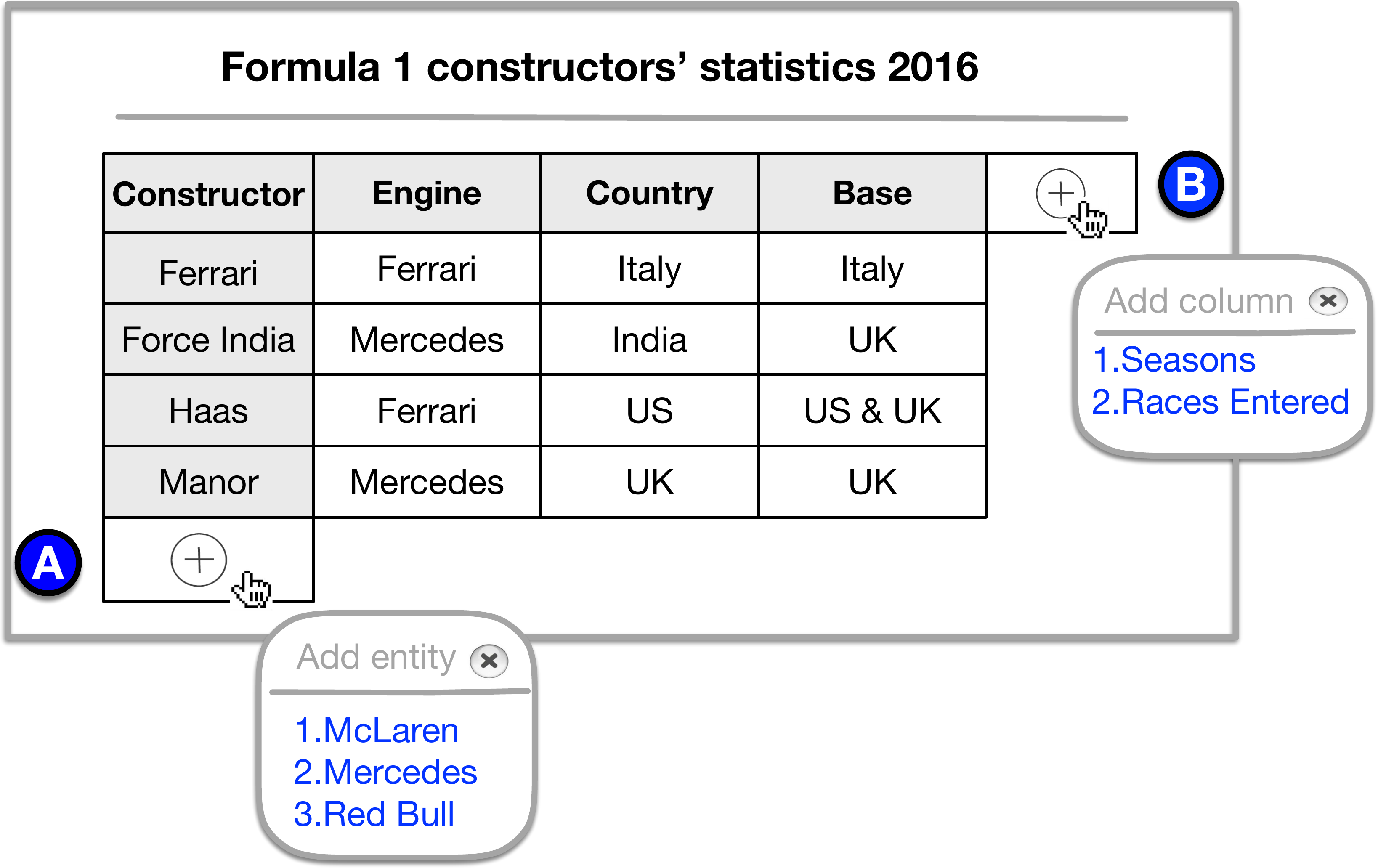} 
   \caption{Envisioned user interface.  Column headings and the leftmost column are marked with a grey background.  The user can populate the table with (A) additional entities (rows) and (B) additional column headings.  The suggestions in the pop-ups are updated dynamically as the content of the table changes.}
   \label{fig:flow}
\end{figure}

The scenario we consider in this paper is the following.
We assume a user, working with a table, at some intermediate stage in the process.  At this point, she has already set the caption of the table and entered some data into the table.  
The table is assumed to have a column header (located above the first content row), which identifies each column with a unique label. 
We further narrow the focus of our study to tables with an \emph{entity focus}.  It means that the leftmost column of the table contains entities. 
This can also be imagined as having a designated row heading, which may contain only (unique) entities.  An entity in the context of this work is a specific object with a unique identifier. (We shall show later in the paper, in \S\ref{sec:expdesign:eft}, that a significant portion of tables have an entity focus.)
Against this setting, our objective is to aid the user by offering ``smart suggestions,'' that is, recommending (i) additional entities (rows) and (ii) additional column headings, to be added to the table.  We shall refer to these tasks as \emph{row population} and \emph{column population}, respectively.
See Figure~\ref{fig:flow} for an illustration.  

Let us point out here that some elements of these tasks have been addressed in prior work.  Our work, however, has not only a different overall motivation, but the specific tasks we tackle have not been addressed in these flavors before. We also introduce a number of innovative elements on the component level.

The task of row population relates to the task of \emph{entity set expansion}~\cite{DasSarma:2012:FRT,Bron:2013:EBE,Metzger:2014:ASE,He:2011:SEI, Wang:2008:ISE, Wang:2015:CEU}, where a given set of seed entities (examples) is to be completed with additional entities.
We also have a seed set of entities from the leftmost table column.  But, in addition to that, we can also make use of the column heading labels and the caption of the table.  We show in our experiments, that utilizing these can lead to substantial improvements over using only the seed entities.

The second task, column population, shares similarities with the problem of \emph{schema complement}~\cite{DasSarma:2012:FRT,Lehmberg:2015:MSJ,Bhagavatula:2013:MEM,Yakout:2012:IEA}, where a seed table is to be complemented with related tables that can provide additional columns.  Many of these approaches utilize the full table content and also address the task of merging data into the seed table. Here, our focus is only on finding proper column headings, using the same sources as for row population (i.e., leftmost column, header row, and table caption).
We show in our experiments that this task can be performed effectively. 

In summary, this paper makes the following novel contributions:
\begin{itemize}
	\item We introduce and formalize two specific tasks for providing intelligent assistance with tables: row population and column population (\S\ref{sec:overview}).
	\item We present generative probabilistic methods for both tasks, which combine existing approaches from the literature with novel components (\S\ref{sec:rows} and \S\ref{sec:columns}).
	\item We design evaluation methodology and develop a process that simulates a user through the process of populating a table with data (\S\ref{sec:expdesign}).
	\item We perform an experimental evaluation and carry out a detailed analysis of performance (\S\ref{sec:eval_row} and \S\ref{sec:eval_col}).
\end{itemize}
All resources developed within this study are made publicly available at \url{http://bit.ly/sigir2017-table}.
\section{Related Work}
There is a growing body of work on web tables and spreadsheets, addressing a range of tasks, including table extension, table completion, table search, table mining, etc. The task of row population is also related to the problem of entity set completion.

\paragraph{Table Extension/Completion}
Extending a local table with additional columns based on the corpus of tables is a relatively new research area. 
The Mannheim Search Joins Engine~\cite{Lehmberg:2015:MSJ} operates on a corpus of web tables, searches for tabular data describing entities in the local table, and then picks relevant columns from the top-$k$ candidate tables to merge.  With a focus on Wikipedia tables,~\citet{Bhagavatula:2013:MEM} target column-matched tables with the local table and perform correlation mining to find ``interesting'' numeric columns.  
InfoGather~\cite{Yakout:2012:IEA} is a table augmentation framework based on topic sensitive PageRank for matching the local table against web tables. The context surrounding the tables is leveraged in a machine learning framework, where the similarity between two tables is captured via a set of features. 
Related tables can be utilized not only for column extension, but for row extension as well. Methods to detect related tables are proposed for the relatedness capture framework in~\cite{DasSarma:2012:FRT}. Two types of table relatedness are identified: entity complement and schema complement. Entity complement tables can be united to produce a meaningful table, and schema complement tables can provide additional meaningful columns.  
\emph{Table completion} refers to the task of filling missing values in a local table. \citet{Ahmadov:2015:THI} propose a hybrid data imputation approach, relying on the characteristics of missing values, in order to (i) look up missing values from web data, (ii) predict them using machine learning methods, or (iii) combine both to find the most appropriate values.  To look up missing values, two keyword subqueries are created from the input table, to search entities and attributes separately.  These resemble our row and column populating subtasks. However, \citet{Ahmadov:2015:THI} have a different target and merge the two search results for table selection. 
\paragraph{Table Search and Mining}
There has been an increasing research interest in mining and searching table content, see, e.g.,~\cite{Cafarella:2008:WEP,Cafarella:2011:SDW,JM:2009:HDW,Sarawagi:2014:OQQ,Venetis:2011:RST,Zhang:2013:ISM}.
Wikipedia's tables contain rich, semi-structured encyclopedic content that is hard to query. 
\citet{Munoz:2014:ULD} extract factual content from Wikipedia tables in the form of RDF triples, contributing to recovering table semantic and discovering table relations. 
Apart from the factual content extraction from web tables, \emph{table mining} also covers tasks like \emph{table interpretation}~\cite{Munoz:2014:ULD, Cafarella:2008:WEP, Venetis:2011:RST} and \emph{table recognition}~\cite{Zwicklbauer:2013:TDW, Crestan:2011:WTC}.

\citet{Cafarella:2008:WEP} extracted 14.1 billion HTML tables from a Google crawl, estimating that 154 million of them contain high-quality relational data. 
The relations extracted from these represent a valuable data resource. 
To disambiguate web tables,~\citet{Zwicklbauer:2013:TDW} propose a methodology to annotate table headers with semantic type information based on the column's content. Similarly,~\citet{Crestan:2011:WTC} present a supervised framework for classifying HTML tables into their taxonomy.
In addition to factual content and relations, numeric attributes are present in a vast number of web tables. However, web tables are not systematic and cannot be used, e.g., for aggregation. To improve the usability of quantities in heterogenous web tables, a line of work aims at detecting quantity mention~\cite{Dong:2014:KVW, Madaan:2016:NRE,Ibrahim:2016:MSE,Roy:2015:RQN,Sellam:2015:RQD} and canonicalizing table quantities~\cite{Ibrahim:2016:MSE,Bhagavatula:2015:TEL,Limaye:2010:ASW,Mulwad:2013:SMP}. 
Another line of work focuses on extracting and fusing numeric attribute values and numeric expressions in  natural language text~\cite{Dong:2014:KVW, Madaan:2016:NRE,Roy:2015:RQN,Sellam:2015:RQD}.  

%

Tables could be well searched and mined for question answering or for extending knowledge bases.
~\citet{Yin:2016:NEL} investigate the task of executing queries on knowledge base tables using Neural Enquirer, which is a fully neuralized DNNs model, both for query planning and for query execution. 
\citet{Sekhavat:2014:KBA} describe a probabilistic method that augments an exiting knowledge base (YAGO). In~\cite{Wang:2012:UTW}, a table search engine is applied to further expand and enrich Probase, which is a universal probabilistic taxonomy framework capable of understanding the entities, attributes and values in web tables. Knowledge Vault is created as a probabilistic knowledge base~\cite{Dong:2014:KVW} by analyzing the extracted content from tabular data, along with other web resources.

\paragraph{Entity Set Completion}
The problem of row population is related to the task of \emph{set completion} or \emph{list expansion}, which is to generate a ranked list of entities starting from a small set of seed entities~\cite{DasSarma:2012:FRT}. \citet{Bron:2013:EBE} propose an approach that combines structure-based and text-based similarity between a candidate entity and the seed entities. 
The QBEES framework~\cite{Metzger:2014:ASE} is designed as an aspect-based entity model to find similar entities based on one or more example entities. 
\citet{He:2011:SEI} focus on entity list data, by picking the top-$k$ entities based on relevance between the candidate entity and seed entities, and then iteratively ranking them according to a combination of relevance and coherence.  
Similar iterative steps are conducted in~\cite{Wang:2015:CEU,Wang:2008:ISE}.  \citet{Wang:2008:ISE} use a random walk method for ranking during iterations.  \citet{Wang:2015:CEU} focus on web tables, instead of entity lists, with the help of a web table search engine, called WTS. 


\section{Problem Statement}
\label{sec:overview}

In this section, we provide a formal description of the tasks we propose to undertake.
We refer to Table~\ref{tbl:notation} for our notation. 

\begin{mydef}[Table]
	A table $T$ is grid of cells, which hold values, arranged in $n+1$ rows and $m$ columns.  The top row is a special designated place, where the column headings reside.  It is followed by $n$ regular (content) rows.
	We let $L=(l_1,\dots,l_m)$ be the list of column heading labels.  In addition to the grid content, the table also has a caption $c$.
	


\end{mydef}

\begin{mydef}[Entity-Focused Table]
	A table is said to be entity-focused, if its leftmost column contains only entities as values, and those entities are unique within the column.  We let $E = (e_1,\dots ,e_n)$ be the list of entities corresponding to the leftmost table column.  I.e., the table takes the following shape:

\begin{align*}
    T = & \begin{bmatrix} 
    	l_{1} & l_{2} & \dots & l_{m} \\
    	e_{1} & v_{1,2} & \dots & v_{1,m} \\
		e_{2} & v_{2,2} & \dots & v_{2,m}\\
		\vdots & \vdots & \ddots & \vdots\\
		e_{n} & v_{n,2} & \dots & v_{n,m}\\
    \end{bmatrix},
\end{align*}
where $v_{i,j}$ ($i \in [1..n], j \in [2..m]$) denote the cell values.

\end{mydef}

\noindent
Our objective is to provide intelligent assistance for an user who is working on an entity-focused table.  We shall refer to the table that is being edited by the user as \emph{seed table}.
We assume that the seed table has already been given a caption, and contains some heading labels (\emph{seed labels}) in the top row and some entities (\emph{seed entities}) in the leftmost column.  
Note that we do not make any assumptions about the values in the other table cells.  Essentially, the $v_{i,j}$ values are immaterial, therefore, we omit them in the followings.\footnote{We note that the $v_{i,j}$ values may also be utilized for row/column population.  However, this is left for future work.}  
When we talk about a table containing entity $e$, we always mean the leftmost table column.

Our goal is to present suggestions for the user for extending the seed table with (i) additional entities, and (ii) additional column heading labels.  
%
%
%
Both tasks are approached as a ranking problem: given a seed table, generate a ranked list of entities (for row population) or column labels (for column population).  


\begin{mydef}[Row Population]
Row population is the task of generating a ranked list of entities to be added to the leftmost column of a given seed table, as $e_{n+1}$. 
\end{mydef}

\begin{mydef}[Column Population]
Column population is the task of generating a ranked list of column labels to be added to the column headings of a given seed table, as $l_{m+1}$. 
\end{mydef}

\begin{table}[!t]
  \centering
  \caption{Notation used in this paper.}
  \begin{tabular}{ll}
    \toprule
    Symbol & Description \\
    \midrule
    $T$ & Table \\
    $c$ & Table caption \\
    $E$ & Seed entities $E=(e_1, \dots, e_n)$ \\
    $L$ & Seed column labels $L^{(j)}=(l_1, \dots, l_m)$ \\
    \bottomrule
  \end{tabular}
  \label{tbl:notation}
\end{table}







In the following two sections, we present our approaches for row and column population.  
Following prior studies~\cite{Lehmberg:2015:MSJ,Bhagavatula:2013:MEM,Yakout:2012:IEA,Ahmadov:2015:THI,DasSarma:2012:FRT,Yin:2016:NEL,Cafarella:2008:WEP,Venetis:2011:RST,Crestan:2011:WTC,Sekhavat:2014:KBA,Wang:2012:UTW,Wang:2015:CEU,Bhagavatula:2015:TEL}, we rely heavily on the availability of a large table corpus as an external resource (which, in our case, is extracted from Wikipedia).
Additionally, we also exploit information stored about entities in a knowledge base (in our case, DBpedia).  Further specifics about our data sources are provided in \S\ref{sec:expdesign:data}.


\section{Populating Rows}
\label{sec:rows}

In this section, we address problem of row population using a two-step approach.  We assume that a seed table is given, with a list of $n$ seed entities $E$, a list of $m$ seed column labels $L$, and a table caption $c$.  The task is to generate a ranked list of suggestions for entity $e_{n+1}$, which may be added to the seed table as a new row.
First, we identify a set of candidate entities (\S\ref{sec:rows:cand}), and then rank them in a subsequent entity ranking step (\S\ref{sec:rows:ranking}).



\subsection{Candidate Selection}
\label{sec:rows:cand}

We identify candidate entities using two sources: knowledge base (KB) and table corpus (TC).
In the knowledge base, each entity $e$ is described by a set of properties $\mathcal{P}_e$.  We focus on two specific properties: types and categories.  We discuss these notions in the context of DBpedia, but note that all knowledge bases employ some taxonomy of types. 
Types in DBpedia are assigned from a small ontology (the DBpedia Ontology, containing a few hundred classes). Categories originate from Wikipedia; these do not form a strict is-a hierarchy, and may be seen more like ``semantic sets.'' Categories are in the order of several 100K. 
Intuitively, an entity $e$ that has several types or categories overlapping with those of the seed entities represents a good candidate.
Thus, we rank entities based on the overlap of these properties, and then take the top-$k$ ones as the set of candidates:

\begin{equation*}
	\mathit{score}(e,E) = \Big| \mathcal{P}_e \cap \big( \cup_{i=1}^n \mathcal{P}_{e_i} \big ) \Big| ~.
\end{equation*}
%



\noindent
When using the table corpus, we search for tables that contain the seed entities or have a similar caption to that of the seed table.  This can be efficiently performed using existing retrieval methods against an inverted index of tables.  Specifically, we use either the seed table's caption or seed entities as the search query and rank tables using the BM25 retrieval algorithm.


\subsection{Ranking Entities}
\label{sec:rows:ranking}

We introduce a probabilistic formulation and rank candidate entities according to the multi-conditional probability $P(e|E,L,c)$.  By applying Bayes's theorem and making a full independence assumption between table caption, seed entities,  and seed column labels, we factor this probability as follows:
\begin{align}
	P(e|E,L,c) & = \frac{P(E,L,c|e)P(e)}{P(E,L,c)} \nonumber \\	
	& = \frac{P(E|e) P(L|e) P(c|e) P(e)}{P(E) P(L) P(c)} \nonumber \\
	& \propto P(e|E)P(L|e)P(c|e) ~. \label{eq:joint}
\end{align}	
%
In the last step, we rewrote $P(E|e)$ using Bayes' rule (which cancelled out $P(e)$ and $P(E)$).  We further dropped the probabilities $P(L)$ and $P(c)$ from the denominator, since those are the same across all candidate entities and thus do not influence their ranking.
Then, entities are ranked by multiplying (i) the posteriori probability $P(e|E)$ that expresses entity similarity, (ii) the column labels likelihood $P(L|e)$, and (iii) the caption likelihood $P(c|e)$.  The reason for keeping the latter two probabilities conditioned on the candidate entity is that column labels and captions are very short.  In those cases, the candidate entity offers a richer observation.
Below, we discuss the estimation of each of these probabilities.

Note that entities may be ranked using any subset of the components in Eq.~\eqref{eq:joint}.  We explore all possible combinations in our experimental section (\S\ref{sec:eval_row}).  It is our expectation that using all three sources of evidence (seed entities, seed column labels, and table caption) would result in the best performance.

\subsection{Entity Similarity}

The estimation of $P(e|E)$ corresponds to the task of \emph{entity list completion} (also known as \emph{set/concept expansion} or \emph{query by example}): given a small set of seed entities, complement this set with additional entities.  The general idea is to measure the semantic similarity between the candidate entity and the set of seed entities.  
One line of prior work~\cite{Bron:2013:EBE,Metzger:2014:ASE} relies on a knowledge base for establishing this semantic similarity.
Another family of approaches~\cite{DasSarma:2012:FRT,Wang:2008:ISE, Wang:2015:CEU} leverages a large table corpus for collecting co-occurrence statistics.
We combine both these sources in a single model:
\begin{equation}
	P(e|E) = \lambda_E P_{KB}(e|E) + (1-\lambda_E) P_{TC}(e|E) ~,
	\label{eq:PeE}
\end{equation}
where $P_{KB}$ is based on the knowledge base 
and $P_{TC}$ is the estimate based on the table corpus. 

\subsubsection{Estimation Using a Knowledge Base} 
\citet{Bron:2013:EBE} create a structured entity representation for each entity from the RDF triples describing that entity.  The structured representation of an entity is comprised by the set of relations of the entity.  Each relation $r$ is modeled as a pair, by removing the entity itself from the triples. E.g., given the triple \texttt{$\langle$dbr:Japan, dbo:capital, dbr:Tokyo$\rangle$} describing the entity \textsc{Japan}, the corresponding relation becomes \texttt{$($dbo:capital, dbr:Tokyo$)$}.  
We write $\hat{e}$ to denote the structured representation of entity $e$.  Formally, given a set of subject-predicate-object $(s,p,o)$ triples describing the entity (i.e., the entity stands either as subject or object):
\begin{equation*}
	\hat{e} = \{(p,o) : (s=e,p,o) \} \cup \{(s,p) : (s,p,o=e) \} ~.
\end{equation*}
%
Similarly, each seed entity is represented as a set of pairs: $\hat{e}_1, \dots, \hat{e}_n$.  The set of seed entities is modeled as a multinomial probability distribution $\theta_E$ over the set of relations.
The probability $P(e|E)$ is then obtained by considering all relations that appear in the representation of the candidate entity:
\begin{equation*}
	P_{KB}(e|E) = \sum_{r \in \hat{e}} P(r|\theta_E)
	= \sum_{r \in \hat{e}}\frac{\sum_{i=1}^n \mathbb{1}(r,\hat{e}_i)}{|\theta_E|} ~,
\end{equation*}
where $\mathbb{1}(r,\hat{e}_i)$ is a binary indicator function, which is $1$ is $r$ occurs in the representation of $\hat{e}_i$ and is $0$ otherwise.  The denominator is the representation length of the seed entities, i.e., the total number of relations of all seed entities: $|\theta_E| = \sum_{i=1}^n \sum_{r \in \hat{e}_i} \mathbb{1}(r,\hat{e}_i)$.

Instead of using a single model built for the set of seed entities, we also explore an alternative approach by taking the average pairwise similarity between the candidate and seed entities (similar in spirit to~\cite{He:2011:SEI,DasSarma:2012:FRT}):
\begin{equation*}
	P_{KB}(e|E) \propto \frac{1}{n}\sum_{i=1}^n \mathrm{sim}(e,e_i) ~,
\end{equation*}
where $\mathrm{sim}(e,e_i)$ is a similarity function.  We consider two alternatives for this function.
The first is the \emph{Wikipedia Link-based Measure} (WLM)~\cite{Milne:2008:ELM}, which estimates the semantic relatedness between two entities based on other entities they link to:
\begin{equation*}
	\mathrm{sim}_{WLM}(e,e_i)=1-\frac{\log(\max(|\mathcal{L}_e|,|\mathcal{L}_{e_i}|))-\log(|\mathcal{L}_e \cap \mathcal{L}_{e_i}|)}{\log(|\mathcal{E}|-\log(\min(|\mathcal{L}_e|,|\mathcal{L}_{e_i}|)))} ~,
\end{equation*}
where $\mathcal{L}_e$ is the set of outgoing links of $e$ (i.e., entities $e$ links to) and $|\mathcal{E}|$ is the total number of entities in the knowledge base.
The second similarity function is the Jaccard coefficient, based on the overlap between the outgoing links of entities:
\begin{equation*}
	\mathrm{sim}_{Jacc}(e,e_i)= \frac{|\mathcal{L}_e \cap \mathcal{L}_{e_i}|}{|\mathcal{L}_e \cup \mathcal{L}_{e_i}|} ~.
\end{equation*}

\subsubsection{Estimation Using a Table Corpus} 

Another way of establishing the similarity between a candidate entity $e$ and the set of seed entities $E$ is to obtain co-occurrence statistics from a table corpus (as in~\cite{DasSarma:2012:FRT,Ahmadov:2015:THI}).
We employ a maximum likelihood estimator:
\begin{equation*}
	P_{TC}(e|E)=\frac{\#(e,E)}{\#(E)} ~,
\end{equation*}
where $\#(e,E)$ is the number of tables that contain the candidate entity together with all seed entities, and $\#(E)$ is the number of tables that contain all seed entities.
Provided that the table corpus is sufficiently large, we expect this simple method to provide an accurate estimate.

\subsection{Column Labels Likelihood}

For computing $P(L|e)$, we consider the tables from the table corpus where the entity appears in the leftmost column.  We obtain and combine two different estimates.
The first one is the representation of the entity in terms of the words of the column labels, i.e., an unigram language model (LM). 
The second one is a maximum likelihood estimate using exact label matching (EM), i.e., without breaking labels up to words.
We consider each individual label $l$ from the seed column labels and combine the above two estimates using a linear mixture:
\begin{equation*}
	P(L|e)=\sum_{l \in L} \Big( \lambda_L \big(\prod_{t \in l}P_{LM}(t|\theta_{e})\big) +\frac{(1-\lambda_L)}{|L|} P_{EM}(l|e) \Big) ~.
	\label{eq:PLe}
\end{equation*}
The first component is a Dirichlet-smoothed unigram language model, calculated using: 
\begin{equation*}
	P_{LM}(t|\theta_{e})=\frac{tf(t,e)+\mu P(t|\theta)}{|e|+\mu} ~,
\end{equation*}
where $tf(t,e)$ is the total term frequency of $t$ in column heading labels of the tables that include $e$ in their leftmost column.  One may think of it as concatenating all the column heading labels of the tables that include $e$, and then counting how many times $t$ appears in there.  The length of the entity $|e|$ is the sum of all term frequencies for the entity ($|e|=\sum_{t'}tf(t',e)$).  The background language model $P(t|\theta)$ is built from the column heading labels of all tables in the corpus.

The exact label matching probability is estimated using:
\begin{equation*}
	P_{EM}(l|e) = \frac{\#(l,e)}{\#(e)} ~,
\end{equation*}
where $\#(l,e)$ is the number of tables containing both $e$ and $l$, and $\#(e)$ is the total number of tables containing $e$.

\subsection{Caption Likelihood}

To estimate the caption likelihood given an entity, $P(c|e)$, we combine two different term-based entity representations: one from the knowledge base and one from the table corpus.  Formally:
\begin{equation*}
	P(c|e) = \prod_{t \in c} \big( \lambda_c P_{KB}(t|\theta_e) + (1-\lambda_c) P_{TC}(t|e) \big ) ~.
	\label{eq:Pce}
\end{equation*}
The knowledge base entity representation is an unigram language model constructed from the entity's description (specifically, its abstract in DBpedia).  Smoothing is done analogously to the column labels language model, but the components of the formula are computed differently:
\begin{equation*}
	P_{KB}(t|\theta_e) = \frac{tf(t,e)+\mu P(t|\theta)}{|e|+\mu} ~,
\end{equation*}
where $tf(t,e)$ denotes the (raw) term frequency of $t$ in the entity's description, $|e|$ is the length (number of terms) of that description, and $P(t|\theta)$ is a background language model (a maximum likelihood estimate from the descriptions of all entities in the KB).

To construct a term-based representation from the table corpus, we consider the captions of all tables that include entity $e$:
\begin{equation*}
	P_{TC}(t|e) = \frac{\#(t,e)}{\#(e)} ~,
\end{equation*}
where $\#(t,e)$ denotes the number of tables that contain term $t$ in the caption as well as entity $e$ in the leftmost column. The denominator $\#(e)$ is the total number of tables containing $e$.\footnote{We also experimented with constructing a smoothed language model, similar to how it was done for the KB, but that gave inferior results.}

%

\section{Populating Columns}
\label{sec:columns}

In this section, we address the problem of column population using a two-step approach: we identify a set of candidate column heading labels (or \emph{labels}, for short), and then subsequently rank them. 

\subsection{Candidate Selection}
\label{sec:cols}
We use (i) the table caption, (ii) table entities, and (iii) seed column heading labels to search for similar tables.
The searching method is the same as in \S\ref{sec:rows:cand}, i.e., we use BM25 similarity using either of (i)--(iii) to get a ranking of tables from the table corpus.
From these tables, we extract the column heading labels as candidates (excluding the seed column labels). 
When searching is done using the seed column labels as query, our method is equivalent to the FastJoin matcher~\cite{Wang:2014:ESS} 
(which was also adopted in~\cite{Lehmberg:2015:MSJ}). 
%

\subsection{Ranking Column Labels}
\label{sec:cols:cand}
We are interested in estimating the probability $P(l|E,c,L)$, given $j$ seed labels, the table caption, and a set of entities from the rows.  

\subsubsection{Baseline Approach}

\citet{DasSarma:2012:FRT} consider the ``benefits'' of additional columns. 
The benefit of adding $l$ to table $T$ is estimated as follows:
\begin{equation}
	P(l|L) = LB(L,l) = \frac{1}{|L|}\sum_{l_1 \in L}cs(l_1,l_2) ~,
\label{eq:sb}
\end{equation}
where $L$ denotes column labels and $cs$ is the AcsDB~\cite{Cafarella:2008:WEP} (\emph{Attribute Correlation Statistics Database}) schema frequency statistics, which is given in Eq.~\eqref{eq:acsdb}.  It is more effective to derive the benefit measure by considering the co-occurrence of pairs of labels, rather than the entire set of labels~\cite{DasSarma:2012:FRT}. Eq.~\eqref{eq:acsdb} determines the consistency of adding a new label $l_2$ to an existing label $l_1$: 
\begin{equation}
	cs(l_1,l_2)=P(l_2|l_1)=\frac{\#(l_1,l_2)}{\#(l_1)} ~,
\label{eq:acsdb}
\end{equation}
where $\#(l_1,l_2)$ is number of tables containing both $l_{1}$ and $l_{2}$, and $\#(l_1)$ is the number of tables containing $l_{1}$.

\subsubsection{Our Approach}

Instead of estimating this probability directly, we use tables as a bridge.
We search related tables sharing similar caption, labels, or entities with the seed table. Searching tables with only one aspect similarity is thought as a single method, e.g., searching tables with similar caption has the probability of $P(T|c)$.  All these related tables are candidate tables acting as bridges.  Each candidate table is weighted by considering its relevance with each candidate label, denoted as $P(l|T)$.

By applying the law of total probability, we get:

\begin{equation*}
	P(l|E,c,L) = \sum_T P(l|T) P(T|E,c,L) ~,
\label{eq:ptecl}
\end{equation*}
where $P(l|T)$ is the label's likelihood given a candidate table (see \S\ref{sec:columns:PlT}), and $P(T|E,c,L)$ expresses that table's relevance (see \S\ref{sec:columns:table_rel}).

\subsection{Label Likelihood}
\label{sec:columns:PlT}

Label likelihood, $P(l|T)$, may be seen as the importance of label $l$ in a given table $T$.
The simplest way of setting this probability is \emph{uniformly} across the labels of the table: 
\begin{equation*}
    P(l|T) = 
	\begin{cases}
    	1, & \text{if $l$ appears in $T$}\\
    	0,              & \text{otherwise} ~.
	\end{cases}
\end{equation*}
%

\subsection{Table Relevance Estimation}
\label{sec:columns:table_rel}

Table relevance expresses the degree of similarity between a candidate table and the seed table the user is working with.  Tables with higher relevance are preferred. Specifically, we search for tables by considering the  similarity of the set of entities, table caption, and column labels.
The probability of a candidate table is factored as:
%
\begin{equation*}
	P(T|E,c,L) = \frac{P(T|E)P(T|c)P(T|L)}{P(T)^2} ~.
\end{equation*}
Notice that an independence assumption between $E$, $c$, and $L^{(j)}$ was made.  Further, assuming that the prior probability of a table follows a uniform distribution, the denominator can be dropped. The components of this model are detailed below.

\subsubsection{Entity Coverage}
When selecting a candidate table, the coverage of the tables' entity set is a important factor~\cite{DasSarma:2012:FRT,Ahmadov:2015:THI}. We compute the fraction of the seed table's entities covered by candidate table as:
\begin{equation*}
	P(T|E) = \frac{|T_E \cap E|}{|E|} ~.
\label{eq:msje2}
\end{equation*}
We note that the same concept is used in~\cite{DasSarma:2012:FRT}, where it is referred to as \emph{entity coverage}.

\subsubsection{Caption Likelihood}
Having similar captions is a strong indicator that two tables are likely to have similar contents.
An effective way of calculating caption similarity is to use the seed table's caption as a query against a caption index of the table corpus.
We can use any term-based retrieval model (like BM25 or language modeling) for measuring caption similarity:
\begin{equation*}
	P(T|c) \propto \mathrm{sim}(T_c,c) ~.
\end{equation*}
%
%

\subsubsection{Column Labels Likelihood}

Finally, we estimate the column labels likelihood similar to \citet{Lehmberg:2015:MSJ}, who rank  tables according to the number of overlapping labels:
\begin{equation*}
	P(T|L) = \frac{|T_L \cap L|}{|L|} ~.
\label{eq:msje}
\end{equation*}

\section{Experimental design}
\label{sec:expdesign}

We present the data sets we use in our experiments and our evaluation methodology. We develop an approach that simulates a user through the process of populating a seed table with data.

\subsection{Data}
\label{sec:expdesign:data}

We use the WikiTables corpus~\cite{Bhagavatula:2015:TEL}, which contains 1.6M tables extracted from Wikipedia.  
The knowledge base we use is DBpedia (version 2015-10). 
We restrict ourselves to entities which have an abstract (4.6M in total).

We preprocess the tables as follows.  For each cell that contains a hyperlink we check if it points to an entity that is present in DBpedia.  If yes, we use the DBpedia identifier of the linked entity as the cell's content (with redirects resolved); otherwise, we replace the link with the anchor text (i.e., treat it as a string).


\subsection{Entity-Focused Tables}
\label{sec:expdesign:eft}

Recall that we defined an entity-focused table as one that contains only unique entities in its leftmost column (cf.~\S\ref{sec:overview}).
In addition to being an entity-focused table, we require that the table has at least 6 rows and at least additional 3 columns (excluding the entity column).  We introduce these constraints so that we can simulate a real-world scenario with sufficient amount of content.

In Table~\ref{tbl:statistic}, we report statistics based on what percentage of cells in the leftmost column contains entities.  Let us note here that only those entities are recognized that have a corresponding Wikipedia article. Thus, the reported numbers should be treated as lower bound estimates.  
It is clear that many tables have an entity focus. 


\begin{table}	
\caption{Statistics of table corpus. Constraints mean having $>5$ rows and $>3$ columns.}
\label{tbl:statistic}
\miniskip
\centering
\begin{tabular}{l c r}
\toprule
leftmost column &\# tables&\# tables with \\
(X\% are entities)& total & constraints \\
\midrule
Existing entity&726913&212923\\
60\%&556644&139572\\
80\%&483665&119166\\
100\%&425236&78611\\
100\% unique&376213&53354\\	
\bottomrule
\end{tabular}
\end{table}

\begin{figure}[t]
   \centering
   \includegraphics[width=7.5cm,height=3.5cm]{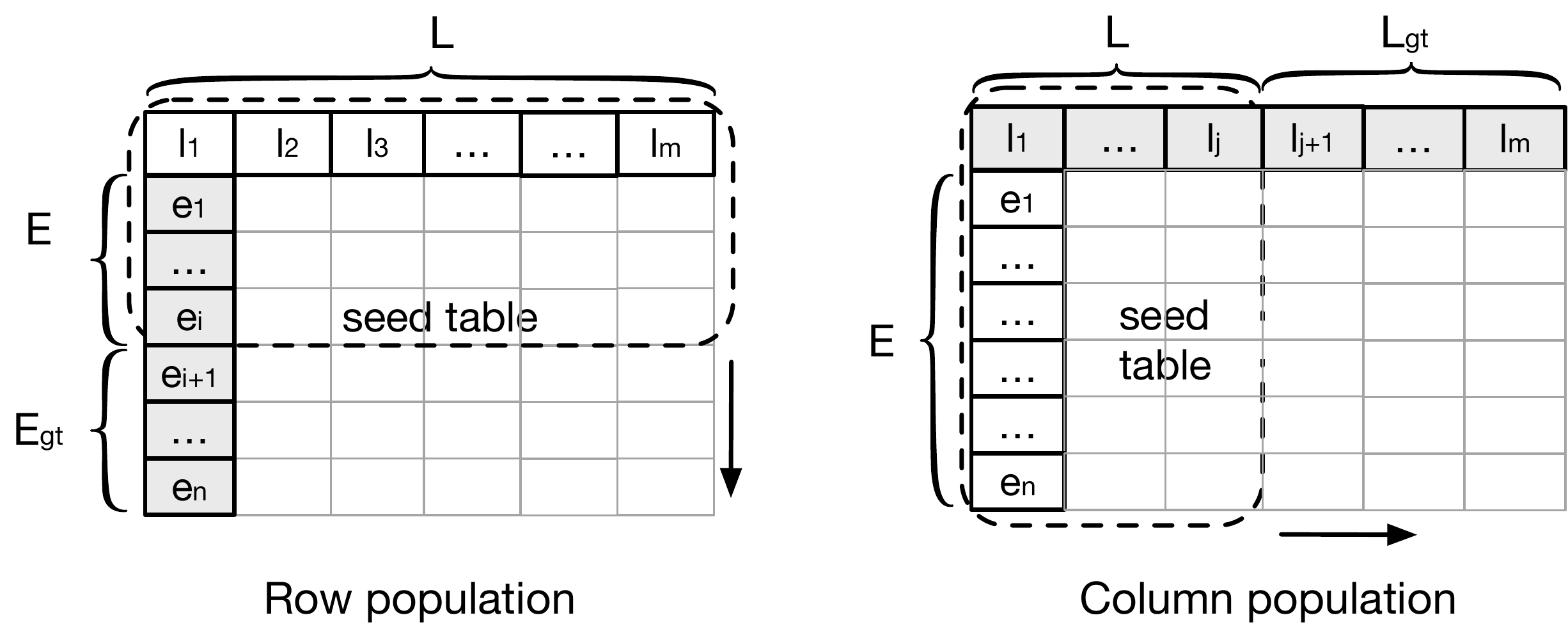} 
   \caption{Illustration of our evaluation methodology.  A part of an existing table is designated as seed table; the entities/column labels outside the seed table serve as ground truth. The arrows indicate the direction of the population.}
   \label{fig:region}
	\shrink
\end{figure}
To be able to perform an automated evaluation without any human intervention, we apply the most strict conditions.  Out of the tables that contain 100\% unique entities in their leftmost column and have at least 6 rows and at least 4 columns (53 K in total), see Table~\ref{tbl:statistic}, we randomly select 1000 tables as our validation set (used for parameter tuning) and another 1000 tables as our test set.  We use a different random selection of validation/test tables for row and column population.
The validation and test sets are excluded from the table corpus during training.
It is important to note that we use all other tables from the corpus when computing statistics, and not only those that classify as entity-focused.

\subsection{Simulation Process}

We evaluate row/column population by starting from an actual (complete) entity-focused table, with $n$ content rows (with an entity in each) and $m$ column headings.  We simulate an user through the process of completing that table by starting with some seed rows/columns and iteratively adding one row/column at a time.  
\begin{itemize}
	\item For evaluating row population, we take entities from the first $i$ rows ($i \in [1..5]$) as seed entities $E$, and use the entities from the remaining rows as ground truth, $\hat{E}$.  We use all column heading labels.
	\item For evaluating column population, we take labels from the first $j$ column ($j \in [1..3]$) as seed column labels $L$, and use the labels from the remaining columns as ground truth, $\hat{L}$.  We use all entities from the table's rows.
\end{itemize}
See Figure~\ref{fig:region} for an illustration.
Notice that we are expanding in a single dimensions at a time; populating both rows and columns at the same time is left for future work.

\subsection{Matching Column Labels}

For the column population task, we are matching string labels (as opposed to unique identifiers).  
Let us consider \texttt{Date} as the ground truth column label.  When the suggested labels are compared against this using strict string matching, then \texttt{date}, \texttt{Dates}, \texttt{date:}, etc. would not be accepted as correct, despite being semantically identical. 
Therefore, we apply some simple normalization steps, on both the ranked and ground truth column labels, before comparing them using strict string matching. 
%
When multiple ranked labels are normalized to the same form, only the one with the highest score is retained.

\subsection{Evaluation Metrics}

Given that the relevance judgments are binary, we use Mean Average Precision (MAP) as our main evaluation metric.  In addition, we also report on Mean Reciprocal Rank (MRR).
We measure statistical significance using a two-tailed paired t-test.
To avoid cluttering the discussion, we report significance testing only for our main metric.

\section{Evaluation of Row Population}
\label{sec:eval_row}

This section presents the evaluation of row population.



\begin{table*}[ht]
  \centering
  \caption{Candidate selection performance for the row population task on the validation set. \#cand refers to the number of candidate entities. Highest recall values are typeset in boldface.}
	\shrink
  \label{tbl:rows:cand}
  \begin{tabular}{ l  rrc  rrc  rrc  rrc  rr }
    \toprule
    & \multicolumn{14}{c}{\#Seed entities ($|E|$)} \\ 
    Method 
    	& \multicolumn{2}{c}{1} & 
    	& \multicolumn{2}{c}{2} & 
    	& \multicolumn{2}{c}{3} & 
    	& \multicolumn{2}{c}{4} & 
    	&\multicolumn{2}{c}{5} \\
    \cline{2-3} \cline{5-6} \cline{8-9} \cline{11-12} \cline{14-15}
    & Recall & \#cand & 
    & Recall & \#cand & 
    & Recall & \#cand & 
    & Recall & \#cand & 
    & Recall & \#cand \\
    \midrule
    (A1) Categories ($k$=256)
    	& 0.6470 & 1721 &
    	& 0.6985 & 2772 &
    	& 0.7282 & 3678 &
    	& 0.7476 & 4507 &
    	& 0.7604 & 5224 \\
    (A2) Types ($k$=4096)
    	& 0.0553 & 7703 &
    	& 0.0577 & 8047 &
    	& 0.0585 & 8225 &
    	& 0.0605 & 8419 &
    	& 0.0600 & 8551 \\
    (B) Table caption ($k$=256)
    	& 0.3966 & 987 &
    	& 0.3961 & 987 &
    	& 0.3945 & 987 &
    	& 0.3938 & 987 &
    	& 0.3929 & 987 \\
    (C) Table entities ($k$=256)
    	& 0.6643 & 312 &
    	& 0.7212 & 458 &
    	& 0.7435 & 589 &
    	& 0.7564 & 689 &
    	& 0.7639 & 759 \\
    \midrule
    (B) \& (C) ($k$=256)
    	& 0.7090 & 1250 &
    	& 0.7464 & 1383 &
    	& 0.7626 & 1505 &
    	& 0.7732 & 1599 &
    	& 0.7788 & 1664 \\
    (A1) \& (B) ($k$=256)
    	& 0.7642 & 2671 &
    	& 0.7969 & 3711 &
    	& 0.8157 & 4610 &
    	& 0.8305 & 5434 &
    	& 0.8405 & 6145 \\
    (A1) \& (C) ($k$=256)
    	& 0.8434 & 1962 &
    	& 0.8885 & 3118 &
    	& 0.9038 & 4117 &
    	& 0.9196 & 5014 &
    	& 0.9285 & 5773 \\
    \midrule
    (A1) \& (B) \& (C) ($k$=256) 
    	& 0.8662 & 2880 &
    	& 0.8997 & 4018 &
    	& 0.9154 & 5005 &
    	& 0.9255 & 5894 &
    	& 0.9329 & 6645 \\
    (A1) \& (B) \& (C) ($k$=4096) 
    	& \textbf{0.9576} & 28733 &
    	& \textbf{0.9718}& 40171 &
    	& \textbf{0.9787} & 49478 &
    	& \textbf{0.9811} & 58021 &
    	& \textbf{0.9821} & 65204 \\
    \bottomrule
  \end{tabular}
\end{table*}

\subsection{Candidate Selection}
\label{eval:row:can}

In \S\ref{sec:rows:cand}, we have introduced four individual methods to select candidates: entity category (A1) and entity type (A2) from the knowledge base, and table caption (B) and table entities (C) from the table corpus.
These methods involve a cut-off threshold parameter $k$; the top-$k$ entities are considered as candidates for the subsequent ranking step.  A larger $k$ value typically implies higher recall.  At the same time, each of the candidate entities will need to be scored, which is a computationally expensive operation.
Therefore, we wish to find a setting that ensures high recall, while keeping the number of candidate entities manageably low (to ensure reasonable response time).
We use the validation set to explore a range of $k$ values: $2^6$, $2^8$, $2^{10}$, and $2^{12}$.
For each method, we select the $k$ value that produces the best recall and candidate entity number ratio. 

The results are reported in the top block of Table~\ref{tbl:rows:cand}.
We observe that more seed entities we have, the better recall gets. This is expected behavior.  Out of the two entity properties from the knowledge base, categories and types, categories performs far better.  For types, even with $k=4096$, the recall is still unsatisfactory.  This is because many of the DBpedia entities have no ontology type information assigned to them.  Moreover, ontology types are more general than categories and result in too many candidates.
The best individual method is (C) table entities; it is the most effective (achieves the highest recall) and the most efficient (produces the lowest number of candidates) at the same time.


To further enhance performance, we combine the individual methods.  However, we exclude type (A2) from this combination, because of its low performance. 
We find that all combinations improve over the single methods.  This means that they capture complimentary aspects.  Combining all three methods (A1+B+C) leads to the best overall performance.
The last two lines of Table~\ref{tbl:rows:cand} show the performance of this combination (A1+B+C) using two different $k$ values.  We find that with a high $k$ value (4096), we are able to achieve close to perfect recall.  The number of candidates, however, is a magnitude larger than with a low $k$ (256).  Motivated by efficiency considerations, we decided not to pay this price and chose to use $k=256$, which still gives us very high recall.

\subsection{Entity Ranking}

\begin{figure}[t]
\begin{tabular}{lll}
	\includegraphics[width=0.15\textwidth]{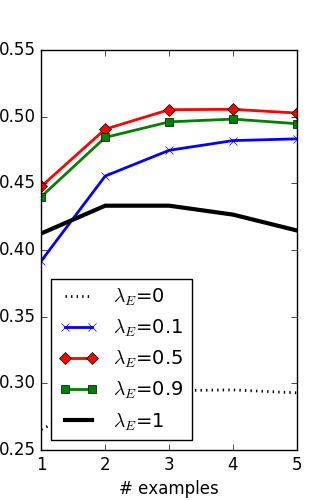}
	&
	\includegraphics[width=0.15\textwidth]{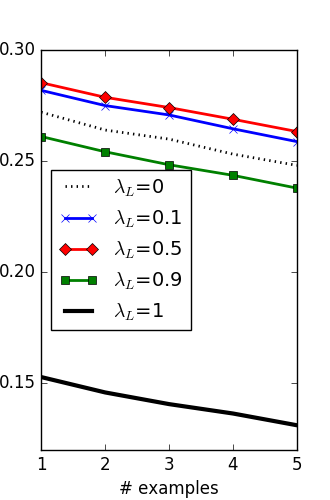}
	&
	\includegraphics[width=0.15\textwidth]{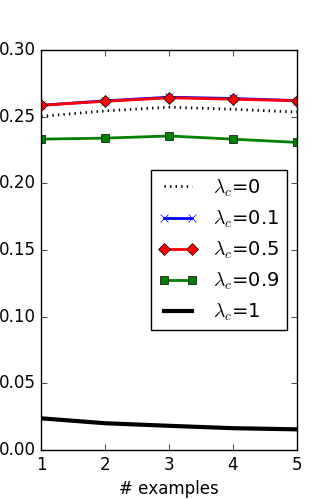}
\end{tabular}
\shrink
\caption{Effect of varying the interpolation parameters for $P(e|E)$ (Left), $P(L|e)$ (Middle), and $P(c|e)$ (Right). The plots show MAP scores measured on the validation set.}
\label{fig:er_lambda}
\end{figure}

Our entity ranking model is comprised of three components:
entity similarity ($P(e|E)$), column labels likelihood ($P(L|e)$), and caption likelihood ($P(c|e)$).
Each of these methods involve an interpolation parameter ($\lambda_E$, $\lambda_L$, and $\lambda_c$, respectively).  We train these parameters on the validation set, by performing a sweep in $0.1$ steps over the $[0..1]$ range.  The effect of varying the parameter values is shown in Figure~\ref{fig:er_lambda}.  It can be seen that the value $0.5$ provides the best setting everywhere.  We also found that there is very little difference in terms of performance when $\lambda$ is in the $0.3..0.7$ range  (hence the choice of showing the $0.1$ and $0.9$ values on the plots).

We start by discussing the performance of individual components, reported in the top block of Table~\ref{tbl:rows:rank}.
The two-component entity similarity model combines estimated based on the knowledge base and the table corpus (cf. Eq.~\eqref{eq:PeE}).  For the former, we compare three alternatives: using relations of entities, as in~\cite{Bron:2013:EBE} (A1), and two similarity methods based on outgoing links of entities: WLM (A2), and Jaccard similarity (A3). 
Out of the three methods, (A1) Relations has the best performance.  However, (A3) has only marginally lower retrieval performance, while being computationally much more efficient.  Therefore, we choose (A3), when it comes to combining it with the other elements of the entity ranking model.
Compared to entity similarity ($P(e|E)$), the other two components (B and C) have much lower performance.  The differences (A3) vs. (B) and (A3) vs. (C) are highly significant ($p < 10^{-5}$).  This means that the knowledge base contributes more.

Next, we combine the individual components to further enhance performance.
The middle block of Table~\ref{tbl:rows:rank} reports results when two components are used.  We find that these combinations improve significantly over the individual methods in all cases ($p < 10^{-5}$).
It is interesting to note that while (C) caption likelihood outperforms (B) column labels likelihood in the individual comparison (significantly so for $\#1..\#3$ seed entities, $p<0.001$), the two perform on a par when combined with (A3) entity similarity.

As expected, using all three component (A3 \& B \& C) results in the best performance.  The differences  between this vs. (A3 \& C) and vs. (B \& C) are significant for any number of seed entities ($p<0.001$); regarding (A3 \& B \& C) vs. (A3 \& B), the differences are significant only for seed entities $\#1$ and $\#5$ ($p<0.05$).  This means that combining information from the knowledge base with column labels from the table corpus yields significant benefits; considering the captions of tables on top of that leads to little additional gain. 

For baseline comparison, we employ the method by~\cite{Bron:2013:EBE}, which combines text-based and structure-based similarity. Note that we used only the structure-based part of their method earlier, as (A1); here, we use their approach in its entirety.  It requires a keyword query, which we set to be the table caption.  
We find that our methods substantially and significantly ($p<10^{-5}$) outperforms this baseline; see the bottom two rows in Table~\ref{tbl:rows:rank}.

One final observation is that performance climbs when moving from a single to two and three seed entities; after that, however, it plateaus.  This behavior is consistent across all methods, including the baseline.  The phenomena is known from prior work~\cite{Bron:2013:EBE,Metzger:2014:ASE,Pantel:2009:WDS}.
 
%


%
\begin{table*}[!t]
  \centering
  \caption{Entity ranking performance on the test set.}
	\shrink
    \begin{tabular}{ l  rrc  rrc  rrc  rrc  rr }
    \toprule
    & \multicolumn{14}{c}{\#Seed entities ($|E|$)} \\ 
    Method 
    	& \multicolumn{2}{c}{1} & 
    	& \multicolumn{2}{c}{2} & 
    	& \multicolumn{2}{c}{3} & 
    	& \multicolumn{2}{c}{4} & 
    	&\multicolumn{2}{c}{5} \\
    \cline{2-3} \cline{5-6} \cline{8-9} \cline{11-12} \cline{14-15}
    & MAP & MRR & 
    & MAP & MRR & 
    & MAP & MRR & 
    & MAP & MRR & 
    & MAP & MRR \\
    \midrule
    (A1) $P(e|E)$ Relations ($\lambda=0.5$)
         & 0.4962& 0.6857&
         & 0.5469& 0.7297&
         & 0.5687& 0.7415&
         & 0.5734& 0.7294&
         & 0.5693& 0.7274\\
    (A2) $P(e|E)$ WLM ($\lambda=0.5$)
	    & 0.4674& 0.6246&
	    & 0.5154& 0.6901&
	    & 0.5293& 0.6930&
	    & 0.5331& 0.6861&
	    & 0.5258& 0.6789\\
    (A3) $P(e|E)$ Jaccard ($\lambda=0.5$)
        & 0.4905& 0.6731&
        & 0.5427& 0.7086&
        & 0.5617& 0.7270&
        & 0.5662& 0.7098&
        & 0.5609& 0.7058\\
    (B) $P(L|e)$
	    & 0.2857& 0.3558&
	    & 0.2878& 0.3518&
	    & 0.2717& 0.3463&
	    & 0.2651& 0.3365&
	    & 0.2585& 0.3336\\
    (C) $P(c|e)$
	    & 0.2348& 0.2656&
	    & 0.2366& 0.2676&
	    & 0.2371& 0.2656&
	    & 0.2350& 0.2614&
	    & 0.2343& 0.2602\\
    \midrule
    (A3) \& (B)
	    & 0.5726& 0.7593&
	    & 0.6108&\textbf{0.8055}&
	    & 0.6189&\textbf{0.7879}&
	    & 0.6182& 0.7755&
	    & 0.6108&\textbf{0.7689}\\
    (A3) \& (C)
	    & 0.5743& 0.7467&
	    & 0.6108& 0.7749&
	    & 0.6221& 0.7746&
	    & 0.6211& 0.7668&
	    & 0.6156& 0.7447\\
    (B) \& (C)
	    & 0.3677& 0.4521&
	    & 0.3715& 0.4508&
	    & 0.3712& 0.4455&
	    & 0.3688& 0.4408&
	    & 0.3667& 0.4378\\
    \midrule
    (A3) \& (B) \& (C)
	    &\textbf{0.5922}&\textbf{0.7729}&
	    &\textbf{0.6260}& 0.8000&
	    &\textbf{0.6339}& 0.7849&
	    &\textbf{0.6348}&\textbf{0.7800}&
	    &\textbf{0.6310}& 0.7630\\
    \emph{Baseline}~\cite{Bron:2013:EBE}
	    & 0.3076& 0.4967&
	    & 0.3273& 0.5156&
	    & 0.3404& 0.5326&
	    & 0.3428& 0.5290&
	    & 0.3406& 0.5202\\
    \bottomrule
  \end{tabular}
  \label{tbl:rows:rank}
\end{table*}


\subsection{Analysis}

\begin{figure}[t]
\center
\includegraphics[scale=0.5]{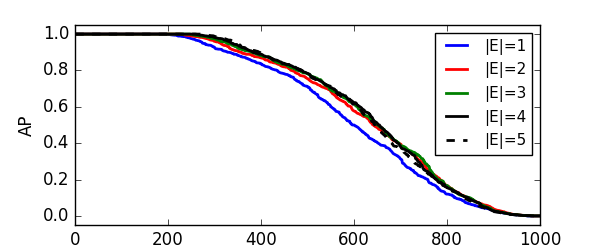}
\shrink
\caption{Performance of individual tables, ordered by decreasing Average Precision, for row population.}
\label{rowap}
\end{figure}

\begin{figure}[t]
\center
\includegraphics[scale=0.4]{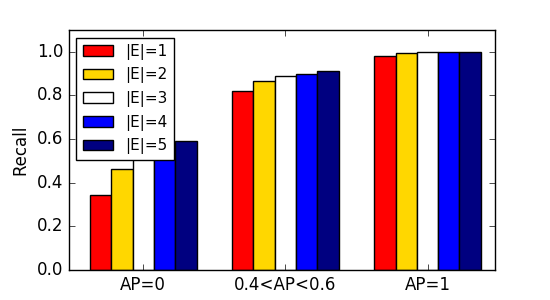}
\shrink
\caption{Recall of candidate selection against entity ranking performance, for row population.}
\label{fg:row:recall}
\end{figure}

Now that we have presented our overall results, we perform further examination on the level of individual tables. 
Figure~\ref{rowap} shows the average precision (AP) scores for the 1000 test tables, ordered by decreasing score. Statistically, there are 285 tables having $AP=1$, 193 tables having $0.4<AP<0.6$, and 42 tables having $AP=0$.  To understand the reasons behind this, we check the recall of the candidate selection step for these three categories; see Figure~\ref{fg:row:recall}. 
In this figure, we can observe that higher recall generally leads to better AP.  Delving deeper, we compute the average number of tables containing at least one ground truth entity, for each of the three groups. When $AP=0$, the number is $18$, for $0.4<AP<0.6$ it is $79$, and for $AP=1$ it is $127$.  It appears that  we could provide excellent suggestions, when there were enough similar tables to the seed table in the table corpus. However, for tables that are ``too unique,'' we would need alternative methods for suggestions.

\section{Evaluation of Column Population}
\label{sec:eval_col}

This section presents the evaluation of column population.
This task relies only on the table corpus; the data set is exactly the same as for row population, see \S\ref{sec:expdesign:data}.

\subsection{Candidate Selection}
In \S\ref{sec:cols}, we have introduced three individual methods to select candidates: table caption (A),  column heading labels (B) and table entities (C). Method (B) actually corresponds to the FastJoin matcher in~\cite{Wang:2014:ESS}. These methods also involve a cut-off threshold parameter $k$, for the same reasons we already discussed in \S\ref{eval:row:can}.
The results are reported in the top block of Table~\ref{tbl:column:cand}. We observe that the more seed labels we have the better recall gets when using labels.  We also explore combinations of pairs of methods as well as using all three. We find that all combinations improve over the single methods, and that combining all three methods leads to the best overall performance.  Our selected method is the second to last in Table~\ref{tbl:column:cand}, motivated by efficiency considerations; for comparison, we also show the performance for $k=4096$.

\begin{table*}[!t]
  \centering
  \caption{Candidate selection performance for the column population task on the validation set.}
	\shrink
  \begin{tabular}{ l  rrc  rrc  rrc  rrc  rr }
    \toprule
    & \multicolumn{8}{c}{\#Seed column labels ($|L|$)} \\ 
    Method 
    	& \multicolumn{2}{c}{1} & 
    	& \multicolumn{2}{c}{2} & 
    	& \multicolumn{2}{c}{3} \\
    \cline{2-3} \cline{5-6} \cline{8-9} 
    & Recall & \#cand & 
    & Recall & \#cand & 
    & Recall & \#cand \\ 

    \midrule    
	(A) Table caption ($k$=256)
	&0.7177&232&
	&0.7115&232&
	&0.7135&231\\
    (B) Column labels ($k$=256)
    &0.2145&115&
    &0.5247&235&
    &0.7014&357\\
    (C) Table entities ($k$=64)
    &0.7617&157&
    &0.7544&156&
    &0.7505&155\\
    \hline
    (A) ($k$=256) \& (B) ($k$=256) \& (C) ($k$=64)
    &0.8799&467&
    &0.8961&572&
    &0.9040&682\\
    (A) ($k$=4096) \& (B) ($k$=4096) \& (C) ($k$=4096)
    &\textbf{0.9211}&2614&
    &\textbf{0.9292}&3309&
    &\textbf{0.9351}&3978\\
    \bottomrule
  \end{tabular}
  \label{tbl:column:cand}
\end{table*}

\subsection{Column Label Ranking}
Our column label ranking model is comprised of two components: table relevance and label likelihood.  
For estimating candidate table relevance, we have three individual methods, using table caption (A), column labels (B), and table entities (C).  All methods use the same estimation of label likelihood (cf.~\S\ref{sec:columns:PlT}). 

We start by discussing the performance of individual methods, which is reported in the top block of Table~\ref{tbl:column:rank}. Of the three, method (C) outperforms the other two, and does significantly so ($p<10^{-5}$). Looking at the tendency of MAP, the increasing number of seed column labels only contributes to method (B). 
When combining two of the methods, all combinations improve significantly over the individual methods ($p<10^{-5}$).  Out of the three, (B) \& (C) performs best in terms of both $MAP$ and $MRR$. 
In the end, putting together all three individual methods delivers the best results.  Also, this combination (A \& B \& C) improves significantly over the combination of any two of the methods ($p<10^{-5}$).

For baseline comparison, we employ the method by~\citet{DasSarma:2012:FRT}. They consider the ``benefits'' of adding additional columns, which expressed in Eq.~\eqref{eq:sb}. We find that our three-component method substantially and significantly ($p<10^{-5}$) outperforms this baseline.  It should be noted that the baseline in~\cite{DasSarma:2012:FRT} uses our candidate selection method to make it comparable; this actually performs better than their original approach.




\begin{table*}[!t]
  \centering
  \caption{Column label ranking performance on the test set.}
	\shrink
    \begin{tabular}{ l  rrc  rrc  rrc  rrc  rr }
    \toprule
    & \multicolumn{8}{c}{\#Seed column labels ($|L|$)} \\ 
    Method 
    	& \multicolumn{2}{c}{1} & 
    	& \multicolumn{2}{c}{2} & 
    	& \multicolumn{2}{c}{3} \\
    \cline{2-3} \cline{5-6} \cline{8-9} 
    & MAP & MRR & 
    & MAP & MRR & 
    & MAP & MRR \\ 

    \midrule 

	(A) Table caption
	&0.2584&0.3496&
	&0.2404&0.2927&
	&0.2161&0.2356\\

    (B) Column labels
    &0.2463&0.3676&
    &0.3145&0.4276&
    &0.3528&0.4246\\
 
    (C) Table entities
    &0.3878&0.4544&
    &03714&0.4187&
    &0.3475&0.3732\\
        
 
    \hline
    (A) \& (B) 
    &0.4824&0.5896&
    &0.4929&0.5837&
    &0.4826&0.5351\\
    (A) \& (C) 
    &0.5032&0.5941&
    &0.4909&0.5601&
    &0.4724&0.5132\\
    (B) \& (C) 
    &0.5060&0.5954&
    &0.5410&0.6178&
    &0.5323&0.5802\\
    \hline
    (A) \& (B) \& (C) 
    & \textbf{0.5863} & \textbf{0.6854} &
    & \textbf{0.5847} & \textbf{0.6690} &
    & \textbf{0.5696} & \textbf{0.6201} \\
    $Baseline$~\cite{DasSarma:2012:FRT}
    &0.4413&0.5473&
    &0.4640&0.5535&
    &0.4535&0.5079\\
    \bottomrule
  \end{tabular}
  \label{tbl:column:rank}
\end{table*}
\begin{figure}[t]
\center
\includegraphics[scale=0.5]{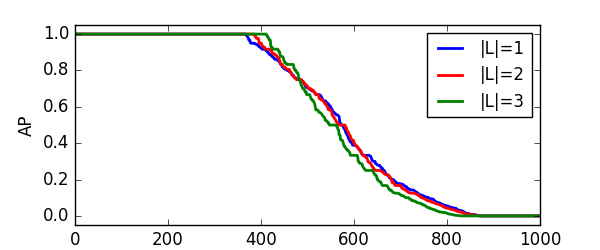}
   \shrink
\caption{Performance of individual tables, ordered by decreasing Average Precision, for column population.}
\label{columnap}
\end{figure}

\begin{figure}[t]
\center
\includegraphics[scale=0.4]{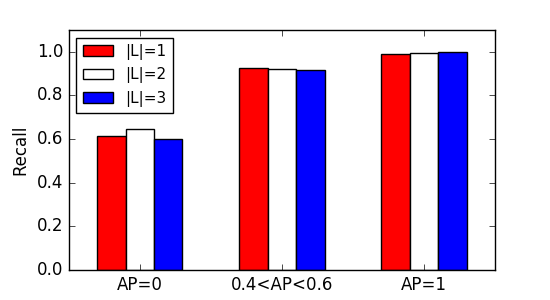}
   \shrink
\caption{Recall of candidate selection against column label ranking performance, for column population.}
\label{fg:col:recall}
\end{figure}

\subsection{Analysis}

Figure~\ref{columnap} plots the performance of individual (test) tables, in decreasing order of average precision score.  We find that there are 427 tables having $AP=1$, 122 tables having $0.4<AP<0.6$, and 186 tables having $AP=0$.  We examine these three table groups, based on performance, in terms of their corresponding recall values from the candidate selection step.  Figure~\ref{fg:col:recall} shows these values (averaged over all tables that fall in the given performance group).  Looking at the number of tables containing at least one ground truth column heading label, it is 204 for $AP=0$, 403 for $0.4<AP<0.6$, and 1114 for $AP=1$. We can draw similar conclusions here as we did for entity ranking.

\section{Conclusion}

In this paper, we have introduced the idea of a smart table assistant and have taken the first steps towards its realization.  Specifically, we have concentrated on tables with an entity focus, and investigated the tasks of row population and column population. 
We have devised methods for each task and showed experimentally how the different components all contribute to overall performance.  For evaluation, we have developed a process that simulates a user through her work of populating a table with data.  Our overall results are very promising and substantially outperform existing baselines.

In future work, we plan to extend the capabilities of our assistant to be able to populate data cells as well with values.
Further along the road, we also wish to relax our requirement regarding the entity focus, and make our methods applicable to arbitrary tables.

\bibliographystyle{ACM-Reference-Format}
\bibliography{sigir2017-table}

\end{document}